\documentclass{article}

\newtheorem{theorem}{Theorem}
\newcounter{countL}
\newtheorem{lemma}[countL]{Lemma}

\usepackage{graphics}
\usepackage{array}
\usepackage{multirow,bigdelim}
\usepackage{enumitem}
\usepackage{amsmath}
\usepackage{caption} 
\usepackage{booktabs,tabu}
\usepackage{color}

\newcommand{\ignore}[1]{}

\begin{document}

\title{SLOCC and LU classification of black holes
with eight electric and magnetic charges}
\author{
Dafa Li$^*$
\and{Maggie Cheng$^\dagger$}
\and{Xiangrong Li$^\ddagger$} 
\and{Shuwang Li}$^\dagger$
}

\date{$^*$ Department of Mathematical Sciences, Tsinghua University,
Beijing, 100084, China
\\ $^\dagger$ Department of Applied Mathematics, Illinois Institute of
Technology, Chicago, IL 60616. USA
\\ $^\ddagger$ Department of Mathematics, University of California-Irvine,
Irvine, CA 92697, USA}

\maketitle

\begin{abstract}

In \cite{Linde}, Kallosh and Linde discussed the SLOCC classification of black
holes. However, the criteria for the SLOCC classification of black
holes have not been given. In addition, the LU classification of black holes has not been studied in the past. In this paper we will consider both SLOCC and LU classification of the STU black holes with four integer electric charges $q_{i} $ and four integer magnetic charges $p^{i}$, $i=0,1,2,3$. Two STU black holes with eight charges are considered SLOCC (LU) equivalent if and only if their corresponding states of three qubits are SLOCC (LU) equivalent. Under this definition, we give criteria for the  classification of the eight-charge STU black holes under SLOCC and under LU, respectively. We will study the classification of the black holes via the classification of SLOCC and LU entanglement of three qubits. We then identify a set of black holes corresponding to the state W of three qubits, which is of interest since it has the maximal average von Neumann entropy of entanglement. Via von Neumann entanglement entropy, we partition the STU black holes corresponding to pure states of GHZ SLOCC class into five families under LU.

\end{abstract}



\section{Introduction}
 SLOCC entanglement classificatio  of $n$ qubits were proposed in \cite{Bennett}, however how to classify them was not discussed. Later, via local ranks, the pure states of three qubits were partitioned into six SLOCC equivalence classes: GHZ, W, AB-C, AC-B, BC-A, and A-B-C \cite
{Dur}. The pure states of four qubits have infinite number of SLOCC classes. They were partitioned into nine families under SLOCC in \cite{Verstraete}. In \cite{Hein}, they characterized and quantified the genuine multi-particle
entanglement of graph states in terms of the Schmidt measure and discussed LU equivalence of two graph states. In \cite{Akhound}, they propose a new
classification for the entanglement in graph states based on generalized
concurrence. 

In \cite{Duff}, Duff related quantum information theory to the physics of
stringy black holes by expressing the entropy of the extremal BPS STU black
hole in terms of Cayley's hyperdeterminant of the coefficients of a
three-qubit pure state. Since then, the correspondence between
 qubits in quantum information theory and black holes in
string theory has been intensively studied \cite{Duff, Duff08, Linde,
Kallosh, Levay, Levay07, Duff-review, Levay10, Borsten, Bellucci, Gabor, Dietrich}.

Kallosh and Linde investigated the relationship between the 3-tangle of
three-qubit states and the entropy of the STU black holes \cite{Linde,
Kallosh}. They showed that the entropy of the axion-dilaton extremal black
hole is related to the concurrence of a 2-qubit state. L\'{e}vay \cite{Levay}
obtained a geometric classification of STU black holes described in the
language of twistors, and established a connection between the black hole
entropy and the average real entanglement of formation. The correspondence between black holes and qubits has also been used to derive the classification of qubits from the knowledge of black holes. Recently, Borsten et al. \cite{Borsten} derived the SLOCC classification of four-qubit
entanglement using the black hole vs qubit correspondence.

Black holes with four non-vanishing
integer charges $q_{0}$, $p^{1}$, $p^{2}$, and $p^{3}$ have been discussed in \cite{Linde, Duff-talk, Levay}. Each of these black holes corresponds to
a state of GHZ SLOCC class for three qubits: 
\begin{equation}
\label{state}
-p^{1}|001\rangle -p^{2}|010\rangle -p^{3}|100\rangle +q_{0}|111\rangle .
\end{equation}

It was indicated in \cite{Duff-talk} that the 4-charge
solution with just $q_{0}$, $p^{1}$, $p^{2}$, and $p^{3}$ may be considered as a bound state of four individual black holes with charges $q_{0}$, $p^{1}$, $p^{2}$, and $p^{3}$, with zero binding energy. 

Clearly, the states in the form of \eqref{state} are real states of GHZ SLOCC class. In 
\cite{Linde}, they indicated that in the case of $q_{0}$ $p^{1}p^{2}p^{3}>0$,
it is related to supersymmetric BPS black holes. For the case of $
p^{1}p^{2}p^{3}<0$, it is related to an extremal non-supersymmetric
non-BPS black holes. {However, the theory of stringy black holes
requires a more detailed classification than the SLOCC classification of three
qubits \cite{Linde}. }In \cite{bh-ghz}, we studied the LU classification of
black holes with four non-vanishing integer charges $q_{0}$, $p^{1}$, $p^{2}$
, and $p^{3}$. Through the LU classification of GHZ SLOCC class, the black
holes are partitioned into seven different families under LU.

In this paper, we will divide the black holes into six families under SLOCC, such
that they have one-to-one correspondence to the six SLOCC classes of three qubits: GHZ, W, AB-C, AC-B, BC-A, and A-B-C. We furthermore partition the black holes corresponding to the GHZ and W
classes under LU.

\section{Motivation}

The general static solution for a spherically symmetric black hole depends
on four integer electric charges, $q_{0}$, $q_{1}$, $q_{2}$, and $q_{3}$,
and four integer magnetic charges, $p^{0}$, $p^{1}$, $p^{2}$, and $p^{3}$ 
\cite{Duff}. Note that $p^{i}$ is not exponential.

The STU black hole entropy $S/\pi $ can be calculated via the 8 charges \cite%
{Behrndt, Linde}. We can simplify the expression for the entropy as $(S/\pi
)^{2}=-\Delta $, where 
\begin{eqnarray}
\Delta &=&(p^{0}q_{0}+p^{1}q_{1}+p^{2}q_{2}-p^{3}q_{3})^{2}  \nonumber \\
&&+4(p^{0}q_{3}-p^{1}p^{2})(p^{3}q_{0}+q_{2}q_{1}),  \label{entropy-1}
\end{eqnarray}

Let $|\Psi \rangle =\sum_{\ell =0}^{7}a_{\ell }|\ell \rangle $ (or $%
\sum_{i,j,k\in \{0,1\}}a_{ijk}|ijk\rangle $) be any pure state of three
qubits A, B and C. Then, we can write the Cayley's hyperdeterminant of $%
|\Psi \rangle $ as follows \cite{DLI-pla}, 
\begin{eqnarray}
\det \Psi &=&(a_{0}a_{7}-a_{1}a_{6}-a_{2}a_{5}+a_{3}a_{4})^{2}  \nonumber \\
&&-4(a_{0}a_{3}-a_{1}a_{2})(a_{4}a_{7}-a_{5}a_{6}).  \label{Cayley}
\end{eqnarray}

It is known that the entanglement measure 3-tangle $\tau _{ABC}=4\;|\det $ $%
\Psi |$ \cite{Miyake, Coffman}. To make a connection between the entropy of
black holes and 3-tangle $\tau _{ABC}$, let $\Delta =\det $ $\Psi $ \cite%
{Duff, Linde, Levay}. To make $\Delta =\det $ $\Psi $, a dictionary between
the eight charges for a black hole and the eight coefficients of a
three-qubit pure state is needed. All the 16 dictionaries were given in \cite%
{bh-ghz}, of which three different dictionaries were given in the
literatures \cite{Duff, Linde, Levay}.

In this paper, we will use the following dictionary \cite{Linde},

\begin{table}[!ht]
\caption{ A dictionary}
\label{tbl:dictionary}
\centering
\begin{tabular}{|l|l|l|l|l|l|l|l|l|}
\toprule
charges & $p^{0}$ & $p^{1}$ & $p^{2}$ & $p^{3}$ & $q_{0}$ & $q_{1}$ & $q_{2}$
& $q_{3}$ \\ 
\midrule
coefficients & $a_{000}$ & $-a_{001}$ & $-a_{010}$ & $-a_{100}$ & $a_{111}$
& $a_{110}$ & $a_{101}$ & $a_{011}$ \\ 
\bottomrule
\end{tabular}
\end{table}

Via Table \ref{tbl:dictionary}, a STU black hole with integer charges $q_{i}$ and $p^{i}$, $%
i=0,1,2,3$, corresponds to the following state:%
\begin{equation}
|\omega \rangle =p^{0}|000\rangle -p^{1}|001\rangle -p^{2}|010\rangle
+q_{3}|011\rangle -p^{3}|100\rangle +q_{2}|101\rangle -q_{1}|110\rangle
+q_{0}|111\rangle .  \label{bh-state}
\end{equation}

It is a correspondence between the black holes with eight integer charges
and pure (unnormalized) states with integer coefficients\ of three qubits
under Table \ref{tbl:dictionary}.

It is known that two pure states $|\psi \rangle
=\sum_{i=0}^{7}c_{i}|i\rangle $ and $|\psi ^{\prime }\rangle
=\sum_{i=0}^{7}c_{i}^{\prime }|i\rangle $ are SLOCC\ (LU) equivalent if and
only if there are local invertible (unitary) operators $\mathcal{A, B, C}$
such that \cite{Dur} 
\begin{equation}
|\psi ^{\prime }\rangle =\mathcal{A}\otimes \mathcal{B}\otimes\mathcal{\ C}%
|\psi \rangle .  \label{Dur}
\end{equation}%
Then, it is natural to define that two STU black holes with charges $q_{i}$
and $p^{i}$, $i=0,1,2,3$, are SLOCC (LU) equivalent if and only if their
corresponding states of three qubits are SLOCC (LU) equivalent.

There are two types for black holes, which are large black holes and small
black holes \cite{Linde}. The large black holes with non-vanishing entropy
correspond to states of GHZ SLOCC class and small black holes with vanishing
entropy correspond states of SLOCC classes A-B-C, A-BC, B-AC, C-AB, and W.
Clearly, the correspondence between two types of black holes and six SLOCC
equivalence\ classes of three qubits is not a one-to-one correspondence. It
needs to partition small black holes into five families to get a 1-1
correspondence between six families of black holes and six SLOCC\ classes of
three qubits.

In this paper, the large black holes are referred to GHZ black holes. The
black holes corresponding to states of SLOCC\ classes W (resp. A-BC, B-AC,
C-AB, A-B-C) are referred to W (resp. A-BC, B-AC, C-AB, A-B-C) black holes.
We will give criteria to describe them.

\section{The\ classification of 8-charge STU black holes via SLOCC\
classification of three qubits}

Let $|\psi \rangle =\sum_{i,j,k}c_{ijk}|ijk\rangle $, with $i, j, k \in \{0, 1\}$, be any pure state of
three qubits. Via local ranks, pure states of three qubits are partitioned
into six SLOCC equivalence\ classes GHZ, W, A-BC, B-AC, C-AB, and A-B-C \cite%
{Dur}. Via the simple arithmetic expressions in coefficients of pure states,
where only +, -, and * operations occur, the necessary and sufficient
condition for each SLOCC\ class was given in Table 1 of \cite{DLI-pla}. Later
on, the similar conditions were given in \cite{Borsten-09}.

The 3-tangle $\tau _{ABC}$ can be written as $\tau _{ABC}=4|D|$, where $D$
is the hyperdeterminant of the coefficients of $|\psi \rangle $.\ In light
of \cite{DLI-pla}, $D$ can be written as 
\begin{eqnarray}
D
&=&(c_{0}c_{7}+c_{1}c_{6}-c_{2}c_{5}-c_{3}c_{4})^{2}-4(c_{2}c_{4}-c_{0}c_{6})(c_{3}c_{5}-c_{1}c_{7})
\label{D-1} \\
&=&(c_{0}c_{7}-c_{1}c_{6}+c_{2}c_{5}-c_{3}c_{4})^{2}-4(c_{1}c_{4}-c_{0}c_{5})(c_{3}c_{6}-c_{2}c_{7})
\label{D-2} \\
&=&(c_{0}c_{7}-c_{1}c_{6}-c_{2}c_{5}+c_{3}c_{4})^{2}-4(c_{0}c_{3}-c_{1}c_{2})(c_{4}c_{7}-c_{5}c_{6}).
\label{D-3}
\end{eqnarray}

Via $D$ in Eqs. (\ref{D-1}, \ref{D-2}, \ref{D-3}) and from \cite{DLI-pla}, $%
\tau _{ABC}=0$ reduces to the following for the SLOCC\ classes A-BC, B-AC,
A-B-C


\begin{equation}
c_{0}c_{7}+c_{1}c_{6}-c_{2}c_{5}-c_{3}c_{4}=0  \label{cd-1}
\end{equation}%
and the following for the SLOCC\ class C-AB 
\begin{equation}
c_{0}c_{7}-c_{1}c_{6}-c_{2}c_{5}+c_{3}c_{4}=0.  \label{cd-2}
\end{equation}%
Eqs. (\ref{cd-1}, \ref{cd-2}) can be used to reduce the conditions in Table
1 \cite{DLI-pla}. Thus, obtain Table \ref{tbl:condition} in which the necessary and sufficient
condition for each SLOCC class in Table 1 of \cite{DLI-pla} is rewritten.

\begin{table}[!ht]
\caption{The necessary and sufficient condition for each SLOCC class.}
\label{tbl:condition}
\scalebox{0.95}{
\centering
\begin{tabular}{|l|l|l|l|l|}
\toprule
SLOCC class & $\tau _{ABC}=0$ & 
\begin{tabular}{l}
$c_{0}c_{3}=c_{1}c_{2}$ \\ 
$c_{5}c_{6}=c_{4}c_{7}$
\end{tabular}
& 
\begin{tabular}{l}
$c_{1}c_{4}=c_{0}c_{5}$ \\ 
$c_{3}c_{6}=c_{2}c_{7}$
\end{tabular}
& 
\begin{tabular}{l}
$c_{3}c_{5}=c_{1}c_{7}$ \\ 
$c_{2}c_{4}=c_{0}c_{6}$%
\end{tabular}
\\ \midrule
GHZ & N  & - & - &-  \\ \midrule
W & Y & N & N & N \\ \midrule
A-BC & Y (Eq. (\ref{cd-1})) & N & Y & Y \\ \midrule
B-AC & Y (Eq. (\ref{cd-1})) & Y & N & Y \\ \midrule
C-AB & Y (Eq. (\ref{cd-2})) & Y & Y & N \\ \midrule
A-B-C & Y (Eq. (\ref{cd-1}) )& Y & Y & Y \\ \bottomrule
\end{tabular}
}
\end{table}

We can obtain the following $D_{bh}$ from $D$ in Eq. (\ref{D-3}), the
following Eqs. (\ref{cd-3}, \ref{cd-4}) from Eqs. (\ref{cd-1}, \ref{cd-2}),
and Table \ref{tbl:slocc} from Table \ref{tbl:condition} by replacing the coefficients $c_{i}$ with the
corresponding charges in the dictionary in Table \ref{tbl:dictionary}. 
\begin{eqnarray}
D_{bh}
&=&(p^{0}q_{0}+p^{1}q_{1}+p^{2}q_{2}-q_{3}p^{3})^{2}+4(p^{0}q_{3}-p^{1}p^{2})(p^{3}q_{0}+q_{2}q_{1}),
\label{tau-bh} \\
\tau _{ABC}^{bh} &=&4|D_{bh}|.
\end{eqnarray}


\begin{eqnarray}
p^{0}q_{0}-p^{1}q_{1}+p^{2}q_{2}+p^{3}q_{3} & =0   \label{cd-3} \\
p^{0}q_{0}+p^{1}q_{1}+p^{2}q_{2}-p^{3}q_{3} & =0    \label{cd-4}
\end{eqnarray}

Thus, black holes with 8 charges $q_{i}$ and $p^{i}$, $i=0,1,2,3$, are
completely partitioned into six families in Table \ref{tbl:slocc} via SLOCC\
classification of three qubits.

We have the following theorem.

\begin{theorem}
A STU black hole with four integer electric charges $q_{i}$, $i=0,1,2,3$,
and four integer magnetic charges $p^{i}$, $i=0,1,2,3$, belongs to one of the six families based on the value of $\tau _{ABC}^{bh}$ as well as conditions A-C: 

\begin{itemize}
\centering
\item[A)] $p^{0}q_{3}=p^{1}p^{2}$ and $q_{2}q_{1}=p^{3}q_{0}$;

\item[B)] $p^{1}p^{3}=p^{0}q_{2}$   and $q_{3}q_{1}=-p^{2}q_{0}$;

\item[C)] $q_{3}q_{2}=-p^{1}q_{0}$  and $p^{2}p^{3}=p^{0}q_{1}$.

\end{itemize}

\begin{itemize}

\item[1).] It belongs to the GHZ black hole family if and only if $\tau _{ABC}^{bh}\neq 0$;

\item[2).] It belongs to the W black hole family if and only if $\tau _{ABC}^{bh}=0$ and it does not
satisfy any of the conditions A, B, or C;

\item[3).] It belongs to the A-BC black hole family if and only if Eq. (\ref{cd-3}) holds and it
satisfies conditions B and C but not A;

\item[4).] It belongs to the B-AC black hole family if and only if Eq. (\ref{cd-3}) holds and it
satisfies conditions A and C but not B;

\item[5).] It belongs to the C-AB black hole family if and only if Eq. (\ref{cd-4}) holds and it
satisfies conditions A and B but not C;

\item[6).] It belongs to the A-B-C black hole family if and only if Eq. (\ref{cd-3}) holds and it
satisfies all conditions A, B and C.

\end{itemize}

\end{theorem}

\begin{table}[!ht]
\caption{Classification of black holes under SLOCC.}
\label{tbl:slocc}
\centering
\begin{tabular}{|l|l|l|l|l|}
\toprule
\multirow{2}{*}{black hole} & \multirow{2}{*}{ $\tau _{ABC}^{bh}=0$ } & $p^{0}q_{3}=p^{1}p^{2}$  & $p^{1}p^{3}=p^{0}q_{2}$  & $q_{3}q_{2}=-p^{1}q_{0}$ \\ 
& & $q_{2}q_{1}=p^{3}q_{0}$ & $q_{3}q_{1}=-p^{2}q_{0}$ & $p^{2}p^{3}=p^{0}q_{1}$ \\
 \midrule
GHZ & N & - &-  &-  \\
\midrule
W & Y & N & N & N \\
\midrule
A-BC & Y (Eq. (\ref{cd-3})) & N & Y & Y \\ 
\midrule
B-AC & Y (Eq. (\ref{cd-3})) & Y & N & Y \\ 
\midrule
C-AB & Y (Eq. (\ref{cd-4})) & Y & Y & N \\ 
\midrule
A-B-C & Y (Eq. (\ref{cd-3})) & Y & Y & Y \\ 
\bottomrule
\end{tabular}
\end{table}

Table \ref{tbl:slocc} summarizes the classification criteria of black holes under SLOCC. 

\section{The classification of GHZ black holes under LU}

\subsection{2-tangles for real states}
We first take a look at the correspondence between 2-tangles for real states and 2-tangles for black holes.

Via the formulas for 2-tangles $\tau _{AB}$, $\tau _{AC}$, and $\tau _{BC}$
in \cite{Dli-sli}, a complicated calculation yields that for any real state
of three qubits, 2-tangles $\tau _{AB}$, $\tau _{AC}$, and $\tau _{BC}$ are

\begin{eqnarray}
\tau _{AB} &=&\allowbreak 4\left(
c_{0}c_{6}-c_{2}c_{4}+c_{1}c_{7}-c_{3}c_{5}\right) ^{2}+2D-2|D|,
\label{2-tang-1} \\
\tau _{BC} &=&4\left( c_{0}c_{3}-c_{1}c_{2}+c_{4}c_{7}-c_{5}c_{6}\right)
^{2}+2D-2|D|,  \label{2-tang-2} \\
\tau _{AC} &=&\allowbreak 4\left(
c_{0}c_{5}-c_{1}c_{4}+c_{2}c_{7}-c_{3}c_{6}\right) ^{2}+2D-2|D|.
\label{2-tang-3}
\end{eqnarray}

Under Table \ref{tbl:dictionary}, from Eqs. (\ref{2-tang-1}, \ref{2-tang-2}, \ref{2-tang-3})
one can obtain the corresponding 2-tangles for black holes as follows. 
\begin{eqnarray}
\tau _{AB}^{bh} &=&\allowbreak 4\left(
p^{0}q_{1}-p^{2}p^{3}-p^{1}q_{0}-q_{3}q_{2}\right) ^{2}+2D_{bh}-2|D_{bh}|,
\label{bh-2-ta-1} \\
\tau _{BC}^{bh} &=&4\left(
p^{0}q_{3}-p^{1}p^{2}-p^{3}q_{0}-q_{2}q_{1}\right) ^{2}+2D_{bh}-2|D_{bh}|,
\label{bh-2-ta-2} \\
\tau _{AC}^{bh} &=&\allowbreak 4\left(
p^{0}q_{2}-p^{1}p^{3}-p^{2}q_{0}-q_{3}q_{1}\right) ^{2}+2D_{bh}-2|D_{bh}|.
\label{bh-2-ta-3}
\end{eqnarray}

In this section, we study classification of GHZ black holes via LU
classification of three qubits.

\subsection{LU classification of GHZ SLOCC class via 2-tangles}

\subsubsection{Via vanishing or unvanishing 2-tangles}

It is known that 2-tangles $\tau _{AB}$, $\tau _{AC}$, and $\tau _{BC}$\ of
three qubits A, B, and C are LU invariants and 2-tangles may vanish for some
states of GHZ SLOCC\ class \cite{Dli-sli}. For example, $\tau _{AB}=\tau
_{AC}=\tau _{BC}=0$ for GHZ state $\frac{1}{\sqrt{2}}(|000\rangle
+|111\rangle )$. Therefore, 2-tangles can be used to partition states of GHZ
SLOCC\ class into the following eight families by letting $\tau _{xy}=0$ or $%
\tau _{xy}\neq 0$, where $xy=AB,AC,$ and $BC$.

(1). $\tau _{AB}=$ $\tau _{AC}=\tau _{BC}=0;$(2). $\tau _{AB}=$ $\tau
_{AC}=0 $, but $\tau _{BC}\neq 0;$ (3). $\tau _{AB}=$ $\tau _{BC}=0$, but $%
\tau _{AC}\neq 0;$ (4). $\tau _{AC}=\tau _{BC}=0$, but $\tau _{AB}\neq 0$;
(5). $\tau _{AB}=0$, but $\tau _{AC}\tau _{BC}\neq 0$; (6). $\tau _{AC}=0$,
but $\tau _{AB}\tau _{BC}\neq 0$; (7). $\tau _{BC}=0$, but $\tau _{AB}$ $%
\tau _{AC}\neq 0$; (8). $\tau _{AB}\tau _{AC}\tau _{BC}\neq 0$.

Specially, 2-tangles $\tau _{AB}$, $\tau _{AC}$, and $\tau _{BC}$ in Eqs. (%
\ref{2-tang-1}, \ref{2-tang-2}, \ref{2-tang-3}) can be used to partition
real states of GHZ SLOCC\ class by letting $\tau _{xy}=0$ or $\tau _{xy}\neq
0$.

\subsubsection{Via the equality of two 2-tangles}
\label{sec:eq2tangle}

By checking equality relation of 2-tangles, states of GHZ SLOCC\ class can
be partitioed into five families: (1). $\tau _{AB}=$ $\tau _{AC}=\tau _{BC}$%
; (2). $\tau _{AB}=$ $\tau _{AC}\neq \tau _{BC}$; (3). $\tau _{AB}=$ $\tau
_{BC}\neq \tau _{AC}$; (4). $\tau _{AC}=$ $\tau _{BC}\neq \tau _{AB}$; (5). $%
\tau _{AC}\neq $ $\tau _{BC}\neq \tau _{AB}\neq \tau _{AC}$.

Specially, 2-tangles $\tau _{AB}$, $\tau _{AC}$, and $\tau _{BC}$ in Eqs. (%
\ref{2-tang-1}, \ref{2-tang-2}, \ref{2-tang-3}) can be used to partition
real states of GHZ SLOCC\ class by checking equality relation of 2-tangles.

\subsubsection{
Reduction of equalities for 2-tangles
}
The conditions used in Section \ref{sec:eq2tangle} can be reduced as follows. Let 
\begin{eqnarray}
\Lambda &=&\allowbreak \left(
p^{0}q_{1}-p^{2}p^{3}-p^{1}q_{0}-q_{3}q_{2}\right) ^{2}, \\
\Pi &=&\left( p^{0}q_{3}-p^{1}p^{2}-p^{3}q_{0}-q_{2}q_{1}\right) ^{2}, \\
\Xi &=&\left( p^{0}q_{2}-p^{1}p^{3}-p^{2}q_{0}-q_{3}q_{1}\right) ^{2}.
\end{eqnarray}

Then, 
\begin{eqnarray}
\tau _{AB}^{bh} &=&4\Lambda +2D_{bh}-2|D_{bh}|, \\
\tau _{BC}^{bh} &=&4\Pi +2D_{bh}-2|D_{bh}|, \\
\tau _{AC}^{bh} &=&4\Xi +2D_{bh}-2|D_{bh}|.
\end{eqnarray}

Thus, the five families in Section 4.2.2 are equivalent to the following
five families: (1). $\Lambda =$ $\Xi =\Pi $; (2). $\Lambda =$ $\Xi \neq \Pi $%
; (3). $\Lambda =$ $\Pi \neq \Xi $; (4). $\Xi =$ $\Pi \neq \Lambda $; (5) $%
\Xi \neq $ $\Pi \neq \Lambda \neq \Xi $.

\subsubsection{
Partition GHZ SLOCC class into five families via
von Neumann entanglement entropy
}

In \cite{Dli-sli}, we showed the following relations.
\begin{eqnarray}
S(\rho _{B}) &=&S(\rho _{A})\quad \mathit{iff}\quad \tau _{AC}=\tau _{BC}, \\
S(\rho _{C}) &=&S(\rho _{A})\quad \mathit{iff}\quad \tau _{AB}=\tau _{BC}, \\
S(\rho _{C}) &=&S(\rho _{B})\quad \mathit{iff}\quad \tau _{AB}=\tau _{AC}.
\end{eqnarray}%
Therefore, by checking equality relation of von Neumann entanglement
entropy, states of GHZ SLOCC class can be partitioned into the
following five families under LU, which are equal to the five families in Section
\ref{sec:eq2tangle}, respectively.

(1). $S(\rho _{A})=S(\rho _{B})=S(\rho _{C})$; (2). $S(\rho _{C})=S(\rho
_{B})\neq S(\rho _{A})$; (3). $S(\rho _{C})=S(\rho _{A})\neq S(\rho _{B})$;
(4). $S(\rho _{B})=S(\rho _{A})\neq S(\rho _{C})$; (5). $S(\rho _{A})\neq
S(\rho _{B})\neq S(\rho _{C})\neq S(\rho _{A})$.

Specially, $S(\rho _{A})$, $S(\rho _{B})$, $S(\rho _{C})$ can be used to
partition real states of GHZ SLOCC\ class by checking equality relation of
von Neumann entanglement entropy.

\vspace{0.1in}

{\bf Remark 1.} In light of \cite{Acin00}, by means of LU transformations any state of three
qubits can be transformed into the following Schmidt decomposition.

\begin{equation}
\label{eq:ASD}
|\psi \rangle =\lambda _{0}|000\rangle +\lambda _{1}e^{i\varphi }|100\rangle
+\lambda _{2}|101\rangle +\lambda _{3}|110\rangle +\lambda _{4}|111\rangle ,
\end{equation}%
where $\lambda _{i}\geq 0$, $\sum_{i=0}^{4}\lambda _{i}^{2}=1$, $\phi $ is
referred to as the phase of $|\psi \rangle $, and here $\phi \in \lbrack
0,2\pi )$. Eq. (\ref{eq:ASD}) is referred to Ac\'{\i}n\ et al.'s Schmidt
decomposition (ASD) of $|\psi \rangle $ (\cite{Acin00}).

Note that ASD of a real state of GHZ SLOCC\ class may be real or not. For
example, $\frac{1}{\sqrt{5}}(-|000\rangle -|001\rangle +|010\rangle
+|100\rangle +|111\rangle )$, its ASD is $\frac{1}{\sqrt{10}}(2|000\rangle
+e^{i(\pi /2)}|100\rangle +|101\rangle +|110\rangle +\sqrt{3}|111\rangle )$.

Therefore, we cannot use classification of real ASDs to study classification
of real states of GHZ SLOCC class. Similarly, we cannot discuss
classification of GHZ-black-holes via classification of real ASDs.

\subsection{LU classification of GHZ black holes via 2-tangles}

\subsubsection{Via vanishing or unvanishing 2-tangles}

Black holes with integer charges always correspond to real states of three qubits. Let us study LU classification of GHZ black holes by means of LU classification of real states of GHZ SLOCC\ class. For GHZ\ SLOCC\ class, $D\neq 0$, and for GHZ black holes, $D_{bh}\neq 0$ in Eq. (\ref{tau-bh}). Via Eqs. (\ref{bh-2-ta-1}, \ref{bh-2-ta-2}, \ref{bh-2-ta-3}), we partition GHZ black holes into eight families under LU.

\begin{itemize}
\item[(1).] $\tau_{AB}^{bh}=\tau _{AC}^{bh}=\tau _{BC}^{bh}=0$;
\item[(2).] $\tau_{AB}^{bh}=\tau _{AC}^{bh}=0$, but $\tau _{BC}^{bh}\neq 0$; 
\item[(3).] $\tau_{AB}^{bh}=\tau _{BC}^{bh}=0$, but $\tau _{AC}^{bh}\neq 0$; 
\item[(4).] $\tau_{AC}^{bh}=\tau _{BC}^{bh}=0$, but $\tau _{AB}^{bh}\neq 0$; 
\item[(5).] $\tau_{AB}^{bh}=0$, but $\tau _{AC}^{bh}\tau _{BC}^{bh}\neq 0$; 
\item[(6).] $\tau_{AC}^{bh}=0$, but $\tau _{AB}^{bh}\tau _{BC}^{bh}\neq 0$; 
\item[(7).] $\tau_{BC}^{bh}=0$, but $\tau _{AB}^{bh}$ $\tau _{AC}^{bh}\neq 0$; 
\item[(8).] $\tau_{AB}^{bh}\tau _{AC}^{bh}\tau _{BC}^{bh}\neq 0$.
\end{itemize}

\subsubsection{Via the equality of two 2-tangles}

By checking equality relation of 2-tangles, GHZ black holes can be partitioned into five families: 
\begin{itemize}
\item[(1).] $\tau_{AB}^{bh}=\tau _{AC}^{bh}=\tau_{BC}^{bh}$; 
\item[(2).] $\tau_{AB}^{bh}=\tau _{AC}^{bh}\neq \tau _{BC}^{bh}$;
\item[(3).] $\tau_{AB}^{bh}=\tau _{BC}^{bh}\neq \tau _{AC}^{bh}$; 
\item[(4).] $\tau_{AC}^{bh}=\tau _{BC}^{bh}\neq \tau _{AB}^{bh}$; 
\item[(5).] $\tau_{AC}^{bh}\neq $ $\tau _{BC}^{bh}\neq \tau _{AB}^{bh}\neq \tau _{AC}^{bh}$.
\end{itemize}

\subsubsection{
Partition GHZ black holes into five families via von
Neumann entanglement entropy
}

Von Neumann entanglement entropy has direct impact on our understanding
black holes and is an important mean to describe black holes \cite{mat,
kiran, geo, yi, bia}. For example, a relationship between the Hawking
radiation energy and von Neumann entanglement entropy in a conformal field
emitted by a semiclassical two-dimensional black hole was found \cite{bia}.
It is known that von Neumann entanglement entropy is LU invariant and the
maximal von Neumann entanglement entropy $S(\rho _{x})$ is $\ln 2$, where $%
x=A,B,$ and $C$.

Under Table 1, one can obtain the corresponding von Neumann entanglement
entropy $S^{bh}(\rho _{x})$ for black holes from $S(\rho _{x})$, $x=A,B,C$.
Thus, from Section 2.2.4, we obtain the following five families under LU for
black holes.

(1). $S^{bh}(\rho _{A})=S^{bh}(\rho _{B})=S^{bh}(\rho _{C})$; (2). $%
S^{bh}(\rho _{C})=S^{bh}(\rho _{B})\neq S^{bh}(\rho _{A})$; (3). $%
S^{bh}(\rho _{C})=S^{bh}(\rho _{A})\neq S^{bh}(\rho _{B})$; (4). $%
S^{bh}(\rho _{B})=S^{bh}(\rho _{A})\neq S^{bh}(\rho _{C})$; (5). $%
S^{bh}(\rho _{A})\neq S^{bh}(\rho _{B})\neq S^{bh}(\rho _{C})\neq
S^{bh}(\rho _{A})$.

It is known that $S^{bh}(\rho _{A})=S^{bh}(\rho _{B})=S^{bh}(\rho _{C})=\ln
2 $ if and only if the state is GHZ black hole. Therefore, GHZ black hole
belongs to Family (1). 

\vspace{0.1in}

{\bf Remark 2.} The black holes with four non-vanishing charges $q_{0}$, $p^{1}$, $p^{2}$, and $p^{3}$ were investigated in \cite{Linde, Duff-talk, Levay}. The black holes correspond to the states in
Eq. (\ref{state}) which belong to a subclass of GHZ SLOCC class. The theory of stringy black holes requires a more detailed classification than SLOCC classification of three qubits \cite{Linde}. In \cite{Linde}, the authors partitioned the black holes with four non-vanishing integer charges $q_{0}$, $p^{1}$, $p^{2}$, and $p^{3}$ into two classes: $q_{0}p^{1}p^{2}p^{3}>0$, related to supersymmetric BPS black holes and $q_{0}p^{1}p^{2}p^{3}<0$, related to non-supersymmetric non-BPS black holes. 

In \cite{bh-ghz}, via LU classification of three qubits we partition the black holes with the charges $q_{0}$, $p^{1}$, $p^{2}$, and $p^{3}$ into seven LU inequivalence families. It is trivial to know that LU classification is a more detailed classification than SLOCC classification because each SLOCC class is partitioned into a finite number of LU families.

In \cite{bh-ghz}, we show the BPS\ black holes and non-BPS black
holes are LU equivalent if their only difference is the signs of the
charges. It means that the classification (about BPS and non-BPS) is not LU
or SLOCC classification.

In this paper, we discuss LU classification of all the black holes
which correspond to states of GHZ SLOCC class via the LU classification of
real states of GHZ SLOCC class. While in \cite{bh-ghz}, we only discuss LU
classification of the black holes with four non-vanishing charges $q_{0}$
, $p^{1}$, $p^{2}$, and $p^{3}$.

\section{The classification of W black holes under LU}

\subsection{LU classification of W SLOCC\ class}

\subsubsection{LU classification of pure states of W SLOCC\ class via
2-tangles}

In light of \cite{Dli-sli}, one knows that 2-tangles don't vanish for W
SLOCC\ class. Therefore, we cannot partition W SLOCC\ class by vanishing
some of 2-tangles. On the other hand, we can partition W SLOCC\ class into five families
by checking the equality relation of any two of 2-tangles (see Sec. \ref{sec:LUASD}).

Specially, for real states of W SLOCC class, by\ Eqs. (\ref{2-ta-1}, \ref%
{j-inv-1},\ref{j-1}, \ref{j-2}, \ref{j-3}) in Appendix A, 2-tangles can be
described by $J_{1}^{\prime },J_{2}^{\prime },$ and $J_{3}^{\prime }$ in
Eqs. (\ref{j-1}, \ref{j-2}, \ref{j-3}) in Appendix A. Thus, real pure states
of W SLOCC\ class can be partitioed into five families by checking equality
relation of $J_{1}^{\prime },J_{2}^{\prime },$ and $J_{3}^{\prime }$. Ref.
the first column of Table \ref{tbl:lu}.

\subsubsection{LU classification of ASDs of W SLOCC\ class}

In light of \cite{Acin00, Dli-qip-18, dli-jpa}, by means of LU
transformations any state of W SLOCC\ class can be transformed into the
following Schmidt decomposition.

\begin{equation}
|\psi \rangle =\lambda _{0}|000\rangle +\lambda _{1}e^{i\varphi }|100\rangle
+\lambda _{2}|101\rangle +\lambda _{3}|110\rangle ,  \label{w-lu-0}
\end{equation}%
where $\lambda _{0}\lambda _{2}\lambda _{3}\neq 0$, $\lambda _{i}\geq 0$, $%
\sum_{i=0}^{3}\lambda _{i}^{2}=1$, $\phi $ is referred to as the phase of $%
|\psi \rangle $, and here $\phi \in \lbrack 0,2\pi )$.

For readability, we restate Proposition 1 of \cite{dli-jpa} as follows.

\begin{lemma}
\label{lm:w-lu}
Two pure states of W SLOCC class are LU equivalent if and
only if their ASDs coincide irrespective of the phases.
\end{lemma}


For example, the state W $=\frac{1}{\sqrt{3}}(|001\rangle +|010\rangle
+|100\rangle )$ and the Dicke state $|2,3\rangle $ $=\frac{1}{\sqrt{3}}%
(|011\rangle +|101\rangle +|110\rangle )$ have the same ASD form: $\frac{1}{%
\sqrt{3}}(|000\rangle +|101\rangle +|110\rangle )$. So, W state and $%
|2,3\rangle $ are LU equivalent.

In light of Lemma 1, one can see that any LU equivalence class for the ASDs
of W SLOCC\ class is a singleton irrespective of the phases.

\begin{lemma}
\label{lm:pt}
From Lemma \ref{lm:w-lu} and via Eq. (\ref{w-lu-0}), ASDs of W SLOCC\
class can be partitioned into eight families as follows.

For $\lambda _{1}=0$, but $\lambda _{0}\lambda _{2}\lambda _{3}\neq 0$, we
can partition ASDs of W SLOCC\ class into three families: (1). states with $%
\lambda _{0}=\lambda _{2}=\lambda _{3}$; (2). states with \ $\lambda
_{i}=\lambda _{j}\neq \lambda _{k}$, where $\{i,j,k\}=(0,2,3)$, $(0,3,2)$, $%
(2,3,0)$; (3). states with $\lambda _{i}\neq \lambda _{j}$, $i\neq j$.

For $\lambda _{0}\lambda _{1}\lambda _{2}\lambda _{3}\neq 0$, we can
partition ASDs of W SLOCC\ class into five families:

Family (4) includes the states with $\lambda _{0}=\lambda _{1}=\lambda
_{2}=\lambda _{3}$.

Family (5) includes the states with $\lambda _{i}=\lambda _{j}=\lambda
_{k}\neq \lambda _{l}$, where $%
(i,j,k,l)=(0,1,2,3),(0,1,3,2),(0,2,3,1),(1,2,3,0)$.

Family (6) includes the states with $\lambda _{i}=\lambda _{j}$ \& $\lambda
_{i}\neq \lambda _{k}\neq \lambda _{l}\neq \lambda _{i}$, where $%
(i,j,k,l)=(0,1,2,3)$, $(2,3,0,1)$, $(0,2,1,3)$, $(1,3,0,2)$, $(0,3,1,2)$, $%
(1,2,0,3)$.

Family (7) includes the states $\lambda _{i}=\lambda _{j}\neq \lambda
_{k}=\lambda _{l}$, where $(i,j,k,l)=(0,1,2,3)$, $(0,2,1,3)$, $(0,3,1,2)$.

Family (8) includes the states with $\lambda _{i}\neq \lambda _{j},i\neq j$.
\end{lemma}

\begin{table}[!ht]
\caption{LU classification of W SLOCC\ class and W black holes via 2-tangles}
\label{tbl:lu}
\centering
\begin{tabular}{|l|l|}
\toprule
W SLOCC class & W black holes \\ \hline
$J_{1}^{\prime }=J_{2}^{\prime }=J_{3}^{\prime }$ & $h_{1}=h_{2}=h_{3}$ \\ 
\midrule
$J_{1}^{\prime }=J_{3}^{\prime }\neq J_{2}^{\prime }$ & $h_{1}=h_{3}\neq
h_{2}$ \\ \midrule
$J_{1}^{\prime }=J_{2}^{\prime }\neq J_{3}^{\prime }$ & $h_{1}=h_{2}\neq
h_{3}$ \\ \midrule
$J_{2}^{\prime }=J_{3}^{\prime }\neq J_{1}^{\prime }$ & $h_{2}=h_{3}\neq
h_{1}$ \\ \midrule
$J_{i}^{\prime }\neq J_{j}^{\prime }$, $i\neq j$ & $h_{i}\neq h_{j}$, $i\neq
j$ \\ \bottomrule
\end{tabular}
\end{table}

\subsection{\protect\bigskip The classification of W black holes under LU\ }

\subsubsection{LU classification of W black holes via 2-tangles}

Via Table \ref{tbl:dictionary}, a black hole with the integer charges $q_{i}$ and $p^{i}$, $%
i=0,1,2,3$, corresponds to the state $|\omega \rangle $ in Eq.(\ref{bh-state}%
). Let $\eta ^{\ast }=1/\sqrt{\sum_{i=0}^{3}(q_{i}^{2}+(p^{i})^{2})}$. Then, 
$\eta ^{\ast }|\omega \rangle $ is normal. For $\eta ^{\ast }|\omega \rangle 
$, Eqs. (\ref{bh-2-ta-1}, \ref{bh-2-ta-2}, \ref{bh-2-ta-3}) reduce to 
\begin{eqnarray}
\tau _{BC}^{bh} &=&4\eta ^{\ast 4}h_{1},h_{1}=\left(
p^{0}q_{3}-p^{1}p^{2}-p^{3}q_{0}-q_{2}q_{1}\right) ^{2},  \label{j-1-1} \\
\tau _{AC}^{bh} &=&4\eta ^{\ast 4}h_{2},h_{2}=\left(
p^{0}q_{2}-p^{1}p^{3}-p^{2}q_{0}-q_{3}q_{1}\right) ^{2},  \label{j-2-1} \\
\tau _{AB}^{bh} &=&4\eta ^{\ast 4}h_{3},h_{3}=\left(
p^{0}q_{1}-p^{2}p^{3}-p^{1}q_{0}-q_{3}q_{2}\right) ^{2}.  \label{j-3-1}
\end{eqnarray}%
Note that $h_{1}h_{2}h_{3}\neq 0$\ because 2-tangles don't vanish for W
SLOCC\ class.

Via 2-tangles $\tau _{xy}^{bh}$ and $h_{1},h_{2},$ and $h_{3}$\ in Eqs. (\ref%
{j-1-1}, \ref{j-2-1}, \ref{j-3-1}), the W black holes can be partitioned
into five families in the second column of Table \ref{tbl:lu}.

\subsubsection{LU classification of ASDs of W black holes}
\label{sec:LUASD}

For W black holes, ASD of $\eta ^{\ast }|\omega \rangle $ is of the
following form 
\begin{equation}
\lambda _{0}^{bh}|000\rangle +\delta \lambda _{1}^{bh}|100\rangle +\lambda
_{2}^{bh}|101\rangle +\lambda _{3}^{bh}|110\rangle ,\delta =\pm 1.
\end{equation}%
Then, a calculation yields the following Schmidt decomposition coefficients
of $\eta ^{\ast }|\omega \rangle $\ from Eqs. (\ref{inv-3}-\ref{inv-6}) in
Appendix A

\begin{eqnarray}
\text{ }(\lambda _{0}^{bh})^{2} &=&(\eta ^{\ast })^{2}\rho _{0}^{2}\text{, }%
\rho _{0}^{2}=\frac{\sqrt{h_{1}h_{2}h_{3}}}{h_{1}}=\xi _{0}\sqrt{%
h_{1}h_{2}h_{3}},\xi _{0}=1/h_{1},  \label{bbh-1} \\
(\lambda _{1}^{bh})^{2} &=&(\eta ^{\ast })^{2}\rho _{1}^{2},  \label{bbh-2-}
\\
\rho _{1}^{2} &=&\sum_{i=0}^{3}(q_{i}^{2}+(p^{i})^{2})-\frac{%
h_{2}h_{3}+h_{2}h_{1}+h_{3}h_{1}}{\sqrt{h_{1}h_{2}h_{3}}}=\xi _{1}\sqrt{%
h_{1}h_{2}h_{3}},  \label{bbh-2} \\
\xi _{1} &=&\frac{\rho _{1}^{2}}{\sqrt{h_{1}h_{2}h_{3}}}, \\
(\lambda _{2}^{bh})^{2} &=&(\eta ^{\ast })^{2}\rho _{2}^{2}\text{, }\rho
_{2}^{2}=\frac{\sqrt{h_{1}h_{2}h_{3}}}{h_{3}}=\xi _{2}\sqrt{h_{1}h_{2}h_{3}}%
,\xi _{2}=1/h_{3},  \label{bbh-3} \\
\text{ }(\lambda _{3}^{bh})^{2} &=&(\eta ^{\ast })^{2}\rho _{3}^{2}\text{, }%
\rho _{3}^{2}=\frac{\sqrt{h_{1}h_{2}h_{3}}}{h_{2}}=\xi _{3}\sqrt{%
h_{1}h_{2}h_{3}},\xi _{3}=1/h_{2}.  \label{bbh-4}
\end{eqnarray}

Note that $\rho _{0}\rho _{2}\rho _{3}\neq 0$.


\begin{lemma}
\label{lm:w-bh}
From Lemma \ref{lm:w-lu}, via Schmidt decomposition coefficients in Eqs. (\ref{bbh-1}- \ref{bbh-4}),
the W black holes can be partitioned into eight families as follows.

(i). for the case $\rho _{1}=0$, by means of $\rho _{0},\rho _{2},$ and $%
\rho _{3}$, the W black holes can be partitioned into three families: (1).
states with$\ \rho _{0}=\rho _{2}=\rho _{3}$; (2). states with $\rho
_{0}=\rho _{3}\neq \rho _{2}$, $\ \rho _{0}=\rho _{2}\neq \rho _{3}$, or $%
\rho _{2}=\rho _{3}\neq \rho _{0}$; and (3). states with $\rho _{i}\neq \rho
_{j}$, $i\neq j$.

Note that the above classification can also be obtained by replacing $\rho
_{i}$ with the simpler $h_{i}$ in Eqs. (\ref{bbh-1}- \ref{bbh-4}) and it is
easy to see that the values of $h_{i}$ under $\rho _{1}=0$ are different
from the ones of $h_{i}$ \ in Eqs. (\ref{j-1-1}, \ref{j-2-1}, \ref{j-3-1}).
One can see that the W black holes of which ASDs have $\rho _{1}=0$ are
partitioned into these three families while whole W black holes are divided
into five families in the second column of Table \ref{tbl:lu}.

(ii). for the case $\rho _{1}\neq 0$, by means of $\rho _{i}$ we can
partition W black holes into five families. Note that we can also use the
simpler $\xi _{i}$ in Eqs. (\ref{bbh-1}- \ref{bbh-4})\ to partition
W black holes instead of $\rho _{i}$.

Family (4) includes the states with $\rho _{0}=\rho _{1}=\rho _{2}=\rho _{3}$%
.

Family (5) includes the states with $\rho _{i}=\rho _{j}=\rho _{k}\neq \rho
_{l}$, where $(i,j,k,l)=(0,1,2,3),(0,1,3,2),(0,2,3,1),(1,2,3,0)$.

Family (6) includes the states with $\rho _{i}=\rho _{j}$ \& $\rho _{i}\neq
\rho _{k}\neq \rho _{l}\neq \rho _{i}$,\ where $(i,j,k,l)=(0,1,2,3)$, $%
(2,3,0,1)$, $(0,2,1,3)$, $(1,3,0,2)$, $(0,3,1,2)$, $(1,2,0,3)$.

Family (7) includes the states with $\rho _{i}=\rho _{j}\neq \rho _{k}=\rho
_{l}$, where $(i,j,k,l)=(0,1,2,3)$, $(0,2,1,3)$, $(0,3,1,2)$.

Family (8) includes the states with $\rho _{i}\neq \rho _{j},i\neq j$.
\end{lemma}

Example 1. Kallosh and Linde considered the black hole with non-vanishing $%
p^{1}$, $p^{2}$, $q_{0}$ charges \cite{Linde}. Via Eqs. (\ref{j-1-1}, \ref%
{j-2-1}, \ref{j-3-1}), a calculation yields that $h_{1}=(p^{1}p^{2})^{2}$, $%
h_{2}=(p^{2}q_{0})^{2}$, $h_{3}=(p^{1}q_{0})^{2}$. \ Via Table \ref{tbl:lu}, we can
partition black holes with charges $p^{1}$, $p^{2}$, $q_{0}$ into five
families in Table \ref{tbl:charge} under LU.

\begin{table}[!ht]
\caption{LU\ classification of black holes with charges $p^{1}$, $p^{2}$, and $q_{0}$}
\label{tbl:charge}
\centering
\begin{tabular}{|l|l|}
\toprule
Family & All black holes with \\ \midrule
1 & $|q_{0}|=|p^{1}|=|p^{2}|$ \\ \midrule
2 & $|q_{0}|=|p^{1}|\neq |p^{2}|$ \\ \midrule
3 & $|q_{0}|=|p^{2}|\neq |p^{1}|$ \\ \midrule
4 & $|p^{2}|=|p^{1}|\neq |q_{0}|$ \\ \midrule
5 & $|q_{0}|\neq |p^{1}|$, $|q_{0}|\neq |p^{2}|$, and $|p^{1}|\neq |p^{2}|$.
\\ \bottomrule
\end{tabular}
\end{table}

Example 2. Consider the black holes with non-vanishing $%
p^{0},p^{3},q_{1},q_{2}$. Via Eqs. (\ref{j-1-1}, \ref{j-2-1}, \ref{j-3-1}),
a calculation yields that $h_{1}=(q_{1}q_{2})^{2}$, $h_{2}=(p^{0}q_{2})^{2}$%
, $h_{3}=(p^{0}q_{1})^{2}$. We can partition the black holes by Table \ref{tbl:lu}.
From Eqs. (\ref{bbh-1}-\ref{bbh-4} ), a calculation yields $\rho
_{0}=|p^{0}|,\rho _{1}=|p^{3}|$, $\rho _{2}=|q_{2}|,\rho _{3}=|q_{1}|$. We
can partition the black holes by Lemma 3. The two classifications are
different.

Example 3. Consider the black holes with non-vanishing charges $%
p^{1},p^{2},q_{0},q_{1},q_{2}$, where $p^{1}q_{1}=p^{2}q_{2}$.\ From $%
p^{1}q_{1}=p^{2}q_{2}$, obtain $p^{1}q_{1}^{2}=p^{2}q_{1}q_{2}$ and $%
q_{1}q_{2}=\frac{p^{1}q_{1}^{2}}{p^{2}}$, and then 
\begin{equation}
p^{1}p^{2}+q_{1}q_{2}=\frac{p^{1}}{p^{2}}((p^{2})^{2}+q_{1}^{2}).
\end{equation}

Via Eqs. (\ref{j-1-1}, \ref{j-2-1}, \ref{j-3-1}), we get 
$h_{1}=(p^{1}p^{2}+q_{1}q_{2})^{2}=\left( \frac{p^{1}}{p^{2}}
((p^{2})^{2}+q_{1}^{2})\right) ^{2}$, $h_{2}=(p^{2}q_{0})^{2}$, $h_{3}=(p^{1}q_{0})^{2}$. We can partition the black holes by Table \ref{tbl:lu}.

From Eqs. (\ref{bbh-1}-\ref{bbh-4} ), we obtain $\rho _{0}=\frac{|p^{2}q_{0}|%
}{\sqrt{(p^{2})^{2}+q_{1}^{2}}}$, $\rho _{1}=\frac{|q_{0}q_{1}|}{\sqrt{%
(p^{2})^{2}+q_{1}^{2}}}$, $\rho _{2}=\sqrt{(p^{2})^{2}+q_{1}^{2}}$, $\rho
_{3}=\frac{|p^{1}|}{|p^{2}|}\sqrt{(p^{2})^{2}+q_{1}^{2}}$. We can partition
black holes by Lemma 3. Thus, we get two different classifications.

\vspace{0.1in}

{\bf Remark 3. }The W black holes from examples 1, 2, and 3 are LU inequivalent.

\vspace{0.1in}

{\bf Remark 4. }It is known that the state W has the maximally average von Neumann
entropy of entanglement within W SLOCC\ class \cite{Dli-sli}. The Schmidt
decomposition of W state is $\frac{1}{\sqrt{3}}(|000\rangle +|101\rangle
+|110\rangle )$. From Appendix A, a calculation yields the LU invariants $%
J_{1}=J_{2}=J_{3}=1/9$ and then $\tau _{AB}=\tau _{AC}=\tau _{BC}=4/9$ for W
state. From Eqs. (\ref{j-1-1}, \ref{j-2-1}, \ref{j-3-1}), one can know the
black holes having the maximally average von Neumann entropy of entanglement
within W black holes satisfy the following: 
\begin{eqnarray}
\eta ^{\ast 4}\left( p^{0}q_{3}-p^{1}p^{2}-p^{3}q_{0}-q_{2}q_{1}\right) ^{2}
&=&1/9, \\
\eta ^{\ast 4}\left( p^{0}q_{2}-p^{1}p^{3}-p^{2}q_{0}-q_{3}q_{1}\right) ^{2}
&=&1/9, \\
\eta ^{\ast 4}\left( p^{0}q_{1}-p^{2}p^{3}-p^{1}q_{0}-q_{3}q_{2}\right) ^{2}
&=&1/9.
\end{eqnarray}

\subsubsection{Discussion}
Kallosh and Linde did SLOCC classification of STU black holes
for three qubits, and they indicated that the theory of stringy black holes
requires a more detailed classification than SLOCC classification of three
qubits \cite{Linde}. To this end, we proposed SLOCC classification and LU classification
of STU black holes for three qubits in this paper.

By relating the complex four qubit SLOCC nilpotent orbits to real orbits
(i.e. Kostant-Sekiguchi correspondence), Borsten \cite{Borsten} invoked the
black-hole--qubit correspondence to derive the classification of four-qubit
entanglement. The U-duality orbits resulting from time-like reduction of
string theory from $D=4$ to $D=3$ yield nine entanglement families up to
permutation of the four qubits. Dietrich et al. studied
classification of four-rebit states under SLOCC in \cite{Dietrich} and they divided four-rebit
states into three groups: semisimple, nilpotent and mixed.

If we also include other parameters (not merely the electric and magnetic charges, but NUT charge, warp factor, moduli etc.) of the STU black holes, then LU classification could be of value in the four-qubit classification context.

\section{Summary}

In this paper, we establish criteria to completely partition 8-charge STU black
holes into six classes, which correspond to the six SLOCC\ classes of three
qubits. The criteria are simple arithmetic expressions in the charges. Via
2-tangles, we partition GHZ SLOCC\ class of three qubits and GHZ black holes
under LU. Also, via 2-tangles, we partition W SLOCC\ class and
W black holes into five families under LU. Via ASD, we
partition W SLOCC\ class and W black holes into eight families under LU.
Via von Neumann entanglement entropy, we partition the STU black holes corresponding to pure states of GHZ SLOCC class into five families under LU.

\section*{Acknowledgement}
We thank the reviewers for their profound insights and suggestions.

\section{ Appendix A. ASD of real states of W SLOCC\ class}

For ASD, Acin et al. proposed LU invariants $J_{1}=|\lambda _{1}\lambda
_{4}e^{i\phi }-\lambda _{2}\lambda _{3}|^{2}$, and $J_{i}=(\lambda
_{0}\lambda _{i})^{2}$ for $i=2,3,4$ \cite{Acin00}. It is known that
3-tangle $\tau _{ABC}$ vanishes for W SLOCC\ class. Therefore, $D$ vanishes
for W SLOCC\ class. Let $|\varphi \rangle =\eta \sum_{i=0}^{7}\sigma
_{i}|i\rangle $ be a state of W SLOCC\ class, where $\eta =1/\sqrt{\sum_{i=0}^{7}\sigma _{i}^{2}}$ \;, and $
\sigma _{i}$ is real. Clearly, $|\varphi
\rangle $ is normal. For $|\varphi \rangle $, from Eqs. (\ref{2-tang-1}-\ref%
{2-tang-3}), obtain 
\begin{eqnarray}
\tau _{AB} &=&\allowbreak 4\eta ^{4}\left( \sigma _{0}\sigma _{6}-\sigma
_{2}\sigma _{4}+\sigma _{1}\sigma _{7}-\sigma _{3}\sigma _{5}\right) ^{2},
\label{normal-w-1} \\
\tau _{BC} &=&4\eta ^{4}\left( \sigma _{0}\sigma _{3}-\sigma _{1}\sigma
_{2}+\sigma _{4}\sigma _{7}-\sigma _{5}\sigma _{6}\right) ^{2},
\label{normal-w-2} \\
\tau _{AC} &=&\allowbreak 4\eta ^{4}\left( \sigma _{0}\sigma _{5}-\sigma
_{1}\sigma _{4}+\sigma _{2}\sigma _{7}-\sigma _{3}\sigma _{6}\right) ^{2}.
\label{normal-w-3}
\end{eqnarray}

$|\varphi \rangle $ can be transformed to the following ASD under LU

\begin{equation}
\lambda _{0}|000\rangle \pm \lambda _{1}|100\rangle +\lambda _{2}|101\rangle
+\lambda _{3}|110\rangle .  \label{w-state}
\end{equation}

It is known that $|\varphi \rangle $ is LU equivalent to its ASD. Therefore, 
$|\varphi \rangle $ and its ASD have the same 2-tangles. For the ASD in Eq. (%
\ref{w-state}), Eqs. (\ref{normal-w-1}-\ref{normal-w-3}) reduce to 
\begin{equation}
\tau _{AB}=4(\lambda _{0}\lambda _{3})^{2}, \; \tau _{AC}=4(\lambda _{0}\lambda
_{2})^{2}, \; \tau _{BC}=4(\lambda _{2}\lambda _{3})^{2}.  \label{2-ta-1}
\end{equation}

It is trivial to see that
\begin{equation}
\tau _{AB}=4J_{3},\; \tau _{AC}=4J_{2}, \; \tau _{BC}=4J_{1}.  \label{j-inv-1}
\end{equation}

From Eqs. (\ref{normal-w-1}-\ref{normal-w-3}, \ref{j-inv-1}), we obtain

\begin{equation}
J_{1}=\eta ^{4}J_{1}^{\prime }, \; J_{2}=\eta ^{4}J_{2}^{\prime }, \; J_{3}=\eta^{4}J_{3}^{\prime },
\end{equation}
where 
\begin{eqnarray}
J_{1}^{\prime } &=&\left( \sigma _{0}\sigma _{3}-\sigma _{1}\sigma
_{2}+\sigma _{4}\sigma _{7}-\sigma _{5}\sigma _{6}\right) ^{2},  \label{j-1}
\\
J_{2}^{\prime } &=&\left( \sigma _{0}\sigma _{5}-\sigma _{1}\sigma
_{4}+\sigma _{2}\sigma _{7}-\sigma _{3}\sigma _{6}\right) ^{2},  \label{j-2}
\\
J_{3}^{\prime } &=&\left( \sigma _{0}\sigma _{6}-\sigma _{2}\sigma
_{4}+\sigma _{1}\sigma _{7}-\sigma _{3}\sigma _{5}\right) ^{2}.  \label{j-3}
\end{eqnarray}

From Eqs. (\ref{2-ta-1}, \ref{j-inv-1}), we obtain the following for real states of W SLOCC\
class 
\begin{equation}
J_{1}=\lambda _{2}^{2}\lambda _{3}^{2}, \; J_{2}=\lambda _{0}^{2}\lambda
_{2}^{2}, \; J_{3}=\lambda _{0}^{2}\lambda _{3}^{2}.  \label{inv}
\end{equation}

Since $J_{1}=\lambda _{2}^{2}\lambda _{3}^{2}$, we have $\lambda
_{0}^{4}J_{1}=\lambda _{0}^{4}\lambda _{2}^{2}\lambda _{3}^{2}=J_{2}J_{3}$, from which we obtain 
\begin{equation}
\text{ }\lambda _{0}^{2}=\sqrt{\frac{J_{2}J_{3}}{J_{1}}}\text{. }
\label{inv-3}
\end{equation}

From Eqs. (\ref{inv}, \ref{inv-3}), we obtain 
\begin{eqnarray}
\lambda _{2}^{2} &=&\frac{J_{2}}{\lambda _{0}^{2}}=\sqrt{\frac{J_{2}J_{1}}{%
J_{3}}},  \label{inv-4} \\
\lambda _{3}^{2} &=&\frac{J_{3}}{\lambda _{0}^{2}}=\sqrt{\frac{J_{3}J_{1}}{%
J_{2}}}.  \label{inv-5}
\end{eqnarray}

Finally, from Eqs. (\ref{inv-3}, \ref{inv-4},\ref{inv-5}), we obtain 
\begin{eqnarray}
\lambda _{1}^{2} &=&1-\lambda _{0}^{2}-\lambda _{2}^{2}-\lambda _{3}^{2}=1-%
\frac{J_{2}J_{3}+J_{2}J_{1}+J_{3}J_{1}}{\sqrt{J_{1}J_{2}J_{3}}}  \nonumber \\
&=&1-\frac{J_{2}^{\prime }J_{3}^{\prime }+J_{2}^{\prime }J_{1}^{\prime
}+J_{3}^{\prime }J_{1}^{\prime }}{\left( \sum_{i=0}^{7}\sigma
_{i}^{2}\right) \sqrt{J_{1}^{\prime }J_{2}^{\prime }J_{3}^{\prime }}}.
\label{inv-6}
\end{eqnarray}

\end{document}